# Perspective - On the thermodynamics of perfect unconditional security

Christiana Chamon and Laszlo B. Kish [†]

*Department of Electrical and Computer Engineering, Texas A&M University,
College Station, TX 77843-3128, USA*

**ABSTRACT**

A secure key distribution (exchange) scheme is unconditionally secure if it is unbreakable against arbitrary technological improvements of computing power and/or any development of new algorithms. There are only two families of experimentally realized and tested unconditionally secure key distribution technologies: Quantum Key Distribution (QKD), the base of quantum cryptography, which utilizes quantum physical photonic features; and the Kirchhoff-Law-Johnson-Noise (KLJN) system that is based on classical statistical physics (fluctuation-dissipation theorem).

The focus topic of this paper is the thermodynamical situation of the KLJN system. In all the original works, the proposed KLJN schemes required thermal equilibrium between the devices of the communicating parties to achieve perfect security. However, Vadai, et al, in (Nature) Science Reports 5 (2015) 13653 shows a modified scheme, where there is a non-zero thermal noise energy flow between the parties, yet the system seems to resist all the known attack types. We introduce a new attack type against their system. The new attack utilizes coincidence events between the line current and voltages. We show that there is non-zero information leak toward the Eavesdropper, even under idealized conditions. As soon as the thermal equilibrium is restored, the system becomes perfectly secure again. In conclusion, perfect unconditional security requires thermal equilibrium.

## INTRODUCTION

Secure communication systems[1-4] require a shared secret between the communicating parties. As an example, in Figure 1, the scheme of the symmetric key exchange protocol is shown. Communicating parties A and B (Alice and Bob) use the same secure key $K$ and the same cipher algorithm to encrypt their plaintext messages ($P_A$ and $P_B$). In this way they produce their cyphertexts, which are deterministic functions of their plaintext and the key. The key $K$ is a random bit string, which is a shared secret, and it is used for both encryption and decryption, respectively.

### Secure key exchange

The hard job is the secure key exchange (distribution) part: to share the secret key between Alice and Bob over the communication channel while Eve is monitoring this communication.[3,4] Secure key distribution is already a secure communication when the encryption key is nonexistent yet. The frontiers of the fundamental research of security in communication systems, particularly in the case of information theoretic (unconditional) security (see below), is to find new ways of secure key exchange. A communication protocol cannot be more secure than its key distribution. Therefore, the rest of this paper is focusing on secure key exchange.

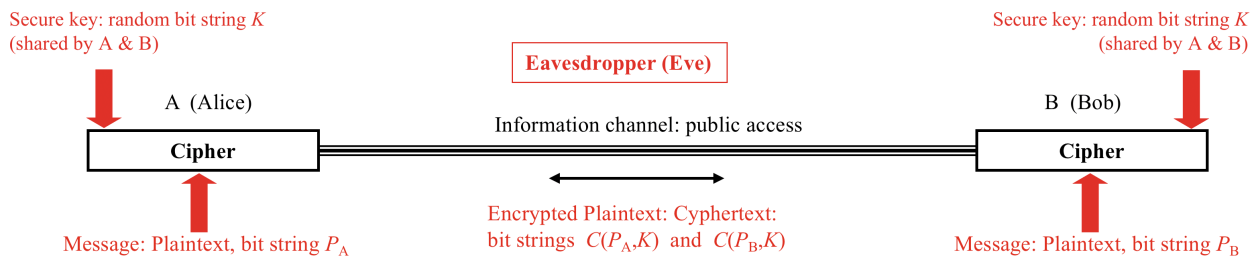

**Figure 1.** Secure communication system with symmetric secure key protocol.

[†] Corresponding author

**Definition of perfect security**

Perfect security of the key exchange (or that of communications) means that Eve's information entropy (that is her uncertainty) about the key (or data) is not decreased by any attack of the communication channel.[1] In other words, the attacks cannot extract any of the secure information.

**Conditional security**

For the complete picture and due to its practical importance, a few words about conditional security. Conditional security means that there is a practical (but not necessarily permanent) condition that limits Eve to extract the data. In these practical cases, the security is typically provided by a *supposedly* hard computational problem that Eve needs to solve to decode the measured data. Such hard problems require an exponential time and/or hardware complexity versus the number of bits to solve. Yet there is no rigorous mathematical proof that these computational problems are indeed hard and that there is no algorithm that can crack the security in polynomial time. Moreover, computational schemes evolve and for example, a quantum computer would be able to efficiently crack most of today's key exchange schemes. Thus conditionally secure schemes are not future-proof. From an information theoretical point of view, they are not secure at all, because Eve has all the necessary data to crack the key by deterministic computations. Even though the computation with the currently available algorithms and technology may take an impractically long time, deterministic computations do not decrease the information entropy. And at the end of the computation, Eve's information entropy about the key will be zero as she will know its bit values. Therefore, conditional security means nonexistent information theoretic security.

**Information theoretic (unconditional) security**

Information theoretically secure (ITS) systems,[2] in the idealized case, hold the information entropy of the key even after an arbitrary attack independently from Eve's conditions and technological advantage including her available computation time. These systems are future proof. In the idealized situation, they represent perfect unconditional security. The security of known ITS systems are based on the laws of physics and cracking the idealized scheme would require breaking some of the relevant laws of physics. There are two families of ITS systems: Quantum Key Distribution and the Kirchhoff-law-Johnson-Noise key exchange.

**Quantum Key Distribution (QKD)**

There are many variations of the QKD scheme (e.g.[3,5-39]). Idealistically, they operate with single photons. The two orthogonal polarization directions represent the values of the two bits. In the original QKD scheme,[5] at the beginning of each clock period, Alice and Bob are randomly switching their polarization bases between the basic direction and an offset of 45 degrees. For Eve, to crack the exchange by measuring, regenerating and forwarding the photons, the knowledge of the actual polarization is missing. Alice and Bob publicly share the directions they used only after the key exchange, which is too late for Eve. Secure key exchange happened when the directions of Alice and Bob were in line. They discard the rest of the measured bits.

For Eve the actual polarization directions, when the bases of Alice and Bob are aligned, represent missing information. For $N$ exchanged key bits, this missing information is $N$ bits thus her information entropy is $N$. Thus to crack a key of $N$ bit length, she would need $N$ bit information that she does not have.

Eve could try to clone the photons, that is, to multiply them without measuring them first. Then they could use the cloned photons for forwarding one in the channel and to test the actual polarization direction with the others. However, this attack, while it could extract some information, would introduce significant bit errors for Alice and Bob that they could detect by testing some parts of the key exchange for errors. The generated bit errors are in accordance with the Quantum No-Cloning Theorem that is the actual law of physics that provides the ultimate security of QKD.

Note, very recently, National Security Agency (NSA) has made a public statement:[40]

"*NSA does not recommend the usage of quantum key distribution and quantum cryptography for securing the transmission of data in National Security Systems (NSS)...*"; "*NSA views quantum-resistant (or post-quantum) cryptography as a more cost effective and easily maintained solution than quantum key distribution.*"

It is true that QKD is still in the fundamental and practical development phase. NSA bases their statement partially on the large body of publications where practical (sometimes already marketed) QKD schemes are cracked (but the related vulnerabilities are usually rectified), (e.g.[6-39]). Yet, we have to note that this statement against QKD is not only unfair but it has some flaws, too.

First of all, QKD (just like the KLJN scheme below) is itself one of the "quantum-resistant (or post-quantum) cryptography" solutions because no quantum computer can crack its key.

Moreover, the NSA statement means by "quantum-resistant (or post-quantum) cryptography" the recent algorithmic key exchange solutions that are resistant against *currently known* quantum computing cracking algorithms. However, due to the lack of mathematical proofs showing that the related math problems are indeed computationally hard, these "quantum-resistant" schemes do not offer unconditional security and their claimed security can evaporate as soon a new quantum or classical algorithm can crack them. The lack of future-proof security also implies the lack of unconditional security[41] of today's algorithm based "quantum-resistant (or post-quantum) cryptography".

Therefore, the above NSA statement against QKD is unfair because it compares apples to oranges.

On the other hand, while we have reservations about the above NSA statement, it inherently brings up the classical physical KLJN for further considerations, see below.

**The Kirchhoff-law-Johnson-Noise (KLJN) key exchange**

The KLJN key distribution[41-94] is a classical statistical physical alternative of QKD for unconditionally secure key exchange. The physical laws of operation are based on Kirchhoff's loop law and the Fluctuation-Dissipation Theorem of statistical physics.[42,43]

The core of the KLJN system operates with 4 resistors (see Figure 2) that form two identical resistor pairs (each pair is of $R_L$ and $R_H$, where $R_L<R_H$) located at communicators A (Alice) and B (Bob), respectively. For each bit exchange period, Alice and Bob randomly and independently choose one of these resistors and connect them to the communication cable (wire).

The key exchanger operates with the thermal noise of the resistors that can be emulated by external Gaussian voltage noise generators. The homogeneous temperature $T$ guarantees that, in the quasi-static frequency limit[4,41,42,89] where the system must operate, the LH (Alice $R_L$, Bob $R_H$) and HL (Alice $R_H$, Bob $R_L$) resistor connections provide the same mean-square voltage and mean-square current in the wire. These are secure levels because the eavesdropper (Eve) knows that the situation is either HL or LH but she is uncertain which one. This is a 1-bit uncertainty (information entropy) for Eve.

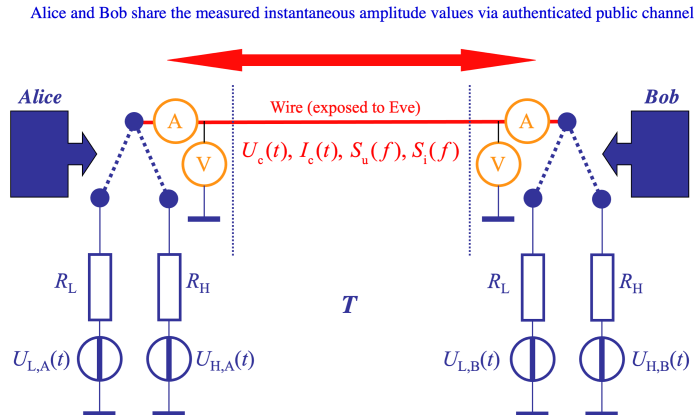

**Figure 2.** The core of the KLJN secure key exchanger scheme. At each end, one of the resistors is randomly selected and connected to the wire. The voltage generators $U_{L,A}$, $U_{H,A}$, $U_{L,B}$, $U_{H,B}$ represent the thermal noise of the resistors or much higher noise temperatures emulated by external, independent, Gaussian noise generators. The homogeneous noise temperature $T$ in the system guarantees that the LH (Alice $R_L$, Bob $R_H$) and HL (Alice $R_H$, Bob $R_L$) resistor connections provide the same mean-square voltage $U_c^2$ and current $I_c^2$, and the same voltage and current noise spectra, $S_u$, $S_i$, in the cable (wire). For defense against various active (invasive) attacks, Alice and Bob are measuring the instantaneous voltage $U_c(t)$ and current $I_c(t)$ amplitudes and exchange these data via an authenticated public channel.

On the other hand, when Alice detects that the noise in the wire is at the secure level, she knows that Bob has connected a different resistor than what she did. Thus, Alice knows Bob's connected resistance value. The same argumentation works for Bob. A random bit generation and secure bit exchange takes place.

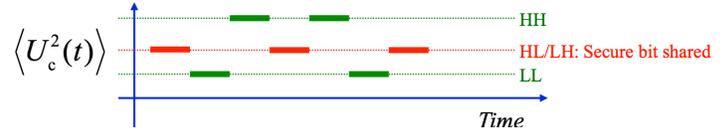

**Figure 3.** Illustration of what Eve sees: the mean-square voltage levels in the wire. The H and L indexes stand for the chosen resistors. The HL and LH levels are identical which makes the bit values corresponding to the HL and LH cases secure. A similar graph exists for the mean-square currents.

Alice and Bob publicly agree on which bit value (0 or 1) the HL situation represents. The LH case represents the opposite bit value.

The unconditional security of the KLJN scheme against passive (listening) attacks is provided by the Second Law of Thermodynamics,[42] which requires thermal equilibrium (homogeneous temperature) for the system.[80,81]

For defense against various active (invasive) attacks, Alice and Bob are measuring the instantaneous voltage and current amplitudes at their sides and exchange these data via an authenticated public channel (Figure 2). In the most advanced schemes, they run a simultaneous cable simulator with these data and compare the results and measured data for consistency.[74]

In the vast body of literature, similarly to the field of QKD, there are attacks, defense techniques, and debates about them. Some of the attacks utilize non-ideal component features that can cause miniscule information leak which can be nullified[75-86] by design and/or privacy amplification.[43,52]

Some attacks are invalid, (e.g.[87]) containing errors in theory/concept, (see[88,89]) and even in their experimental verification, see a critical analysis.[90] Some attacks are conceptually valid (e.g.[91]) but with flaws in design and modeling. They can cause only minor information leak that is removable[92] by design, protocol, and/or privacy amplification.

However, we must note that none of the attacks have be in vain because even the missed ones have contributed to the deeper understanding of the interplay between physics and security.

Finally, we note that there have been attempts to emulate KLJN with proper algorithm, (e.g.[95]) however the situation is disadvantageous compared to a cable because the signals coming from Alice and Bob can be separated, which makes transient based attacks feasible. This situation can be avoided in the KLJN system by a proper choice of noise of bandwidth, line filters, and signal initialization.[41,50]

## SECURITY IN THERMAL EQUILIBRIUM

In the KLJN protocol, the net power flow $P_{AB}$ between Alice and Bob is zero because their resistors have the same (noise) temperature. The noise spectra of the voltage $U_c$ and current $I_c$ in the wire, $S_u$ and $S_i$, respectively, are given by the Johnson formulas of thermal noise:

$$S_u(f) = 4kTR_p, \quad (1)$$

$$S_i(f) = \frac{4kT}{R_s}, \quad (2)$$

where $k$ is the Boltzmann constant, and $R_p$ and $R_s$ are the parallel and serial resultants of the connected resistors, respectively. In the HL and LH cases the resultants are:

$$R_{pLH} = R_{pHL} = \frac{R_L R_H}{R_L + R_H}, \quad (3)$$

$$R_{sLH} = R_{sHL} = R_L + R_H. \quad (4)$$

Equations 1-4 guarantee that the noise spectra and effective voltage and current values in the wire are identical in the LH and HL cases in accordance with the perfect security requirement. In conclusion, the quantities that Eve can access with passive measurements satisfy the following equations that, together with Equations 3-4, form the pillars of security against passive attacks against the KLJN system:

$$U_{LH} = U_{HL}, \quad (5)$$

$$I_{LH} = I_{HL}, \quad (6)$$

$$P_{LH} = P_{HL} = 0. \quad (7)$$

## SECURITY OUT OF EQUILIBRIUM? THE VMG-KLJN SYSTEM

Vadai, Mingesz and Gingl (VMG) has made an impressive generalization attempt:[46] they assumed that the four resistors are different and arbitrarily chosen (with some limitations[58]) and asked the question if the security can be maintained by a proper choice of different temperatures of these resistors.

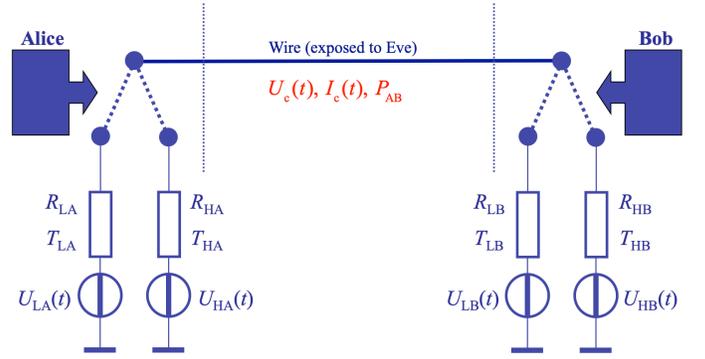

**Figure 4.** The core of the Vadai-Mingesz-Gingl (VMG-KLJN) secure key exchanger scheme. The four resistors are different and they can be freely chosen (though with some limitations because of certain unphysical solutions). One temperature is freely chosen and the other 3 temperatures depend on the resistor values and can be deducted by the VMG-equations 8-10, see Equations 12-14 below.

In search for their solution, they used Equations 5 and 6 and removed Equation 7, the zero power flow condition.

VMG obtained the following solutions for the required mean-square thermal noise voltages of the thermal noise generators of the resistors, where the rms voltage $U_{LA}$ of the resistor $R_{LA}$ is freely chosen:

$$U_{HB}^2 = U_{LA}^2 \frac{R_{LB}(R_{HA} + R_{HB}) - R_{HA}R_{HB} - R_{HB}^2}{R_{LA}^2 + R_{LB}(R_{LA} - R_{HA}) - R_{HA}R_{LA}} = 4kT_{HB}R_{HB}B, \quad (8)$$

$$U_{HA}^2 = U_{LA}^2 \frac{R_{LB}(R_{HA} + R_{HB}) + R_{HA}R_{HB} + R_{HA}^2}{R_{LA}^2 + R_{LB}(R_{LA} + R_{HB}) + R_{HB}R_{LA}} = 4kT_{HA}R_{HA}B, \quad (9)$$

$$U_{LB}^2 = U_{LA}^2 \frac{R_{LB}(R_{HA} - R_{HB}) - R_{HA}R_{HB} + R_{LB}^2}{R_{LA}^2 + R_{LA}(R_{HB} - R_{HA}) - R_{HA}R_{HB}} = 4kT_{LB}R_{LB}B. \quad (10)$$

For more direct statistical physical comparison, we expanded[58] the VMG equations by introducing (on the right hand side) the temperatures of the resistors, where $B$ is the noise bandwidth of the generators, which is identical for all resistors, and the required temperatures of the resistors shown above are determined by the Johnson-Nyquist formula:

$$U^2 = 4kTRB. \quad (11)$$

Thus, from Equations 8-11, the temperatures are:

$$T_{HB} = \frac{R_{LA}}{R_{HB}} T_{LA} \frac{R_{LB}(R_{HA} + R_{HB}) - R_{HA}R_{HB} - R_{HB}^2}{R_{LA}^2 + R_{LB}(R_{LA} - R_{HA}) - R_{HA}R_{LA}}, \quad (12)$$

$$T_{HA} = \frac{R_{LA}}{R_{HA}} T_{LA} \frac{R_{LB}(R_{HA} + R_{HB}) + R_{HA}R_{HB} + R_{HA}^2}{R_{LA}^2 + R_{LB}(R_{LA} + R_{HB}) + R_{HB}R_{LA}}, \quad (13)$$

$$T_{LB} = \frac{R_{LA}}{R_{LB}} T_{LA} \frac{R_{LB}(R_{HA} - R_{HB}) - R_{HA}R_{HB} + R_{LB}^2}{R_{LA}^2 + R_{LA}(R_{HB} - R_{HA}) - R_{HA}R_{HB}}, \quad (14)$$

where $T_{LA}$ is the temperature of resistor $R_{LA}$.

The practical advantage of the VMG-KLJN scheme would appear with inexpensive versions of chip technology where resistance accuracy and its temperature stability are poor.

However, concerning the fundamental physics aspects for security there is a much more important question: What law of physics that guarantees the perfect security of the ideal system?

In the case of the standard KLJN system, that is the *Second Law of Thermodynamics*. However, due to VMG's security claim at nonzero power flow this explanation is seemingly irrelevant in the VMG-KLJN system.

**The FCK1-VMG-KLJN system: different resistors but still in equilibrium**

Recently, Ferdous, Chamon and Kish (FCK)[58] pointed out some excess information leak (compared to classical KLJN protocols) in the VMG-KLJN system under practical conditions. Among others, they proposed three modified VMG-KLJN versions for improvements. One of these schemes, the FCK1-VMG-KLJN scheme[58], is able to operate with 4 different resistors so that during each secure bit exchange period *the connected* resistor pair (one resistor at each side) is in thermal equilibrium, that is, the resistors in the pair have the same temperature. However, the two "secure-choice" resistor arrangements, $R_{HA}\|R_{LB}$ and $R_{LA}\|R_{HB}$ must be at a different temperature, except in the original KLJN scheme where the two resistor pairs (of the HL and LH situations) are identical. (This is a minor security risk but is out of the topic of our present paper).

The condition of zero power is that the geometrical means of the connected resistors in the LH and HL situations are equal.[58] In other words, when we choose three resistors freely, the fourth one is determined by the condition of zero power flow. For example, with chosen $R_{HB}$, $R_{LA}$ and $R_{HA}$, we get:

$$R_{LB} = \frac{R_{HB}R_{LA}}{R_{HA}} \quad (15)$$

Due to the thermal equilibrium during a single bit exchange, the FCK1-VMG-KLJN system has a special role in the study of the non-equilibrium VMG-KLJN protocol in the following section.

In the next section we answer the following question: Is it possible that there is a new, unknown attack that can extract information from the VMG-KLJN system while it is unable to do that with the standard KLJN scheme? If so, the VMG-KLJN arrangement would be just a modified KLJN scheme wthat is *distorted* for a special purpose (free resistor choice) while, as a compromise, its perfect security is given up. It would still have the same foundation of security - the Second Law - but in an imperfect way due to the non-ideality introduced by the nonzero power flow. In the next two sections, we show that indeed this is the case.

## ZERO-CROSSING ATTACK AGAINST THE VMG-KLJN SCHEME

The VMG-KLJN scheme seems to be perfectly secure at non-zero power flow because the voltage and the current are Gaussian processes and their mean-square values are identical in the LH and HL bit situations even though the resistor and related mean-square voltage pairs are different. Therefore, even the power, which is the mean of their product, seems to carry no useful information for Eve. Gaussian processes are perfect information hiders.

Thus we are exploring here a yet uncharted area: the statistics of the coincidence properties of the voltages at the two ends. Whenever the current is zero in the wire, the voltages at the two ends are equal. Let us *sample* the voltages on the wire at these coincidence points: then the voltage in the wire has the same value as that of the generators at Alice and Bob because the current is zero. For an intuitive start, imagine the situation when in the HL case the VMG voltage is very high at the H side and small at the L side while in the LH case the voltages are similar (see Equations 8-10). The Gaussian process is statistically confined to the order of the RMS value thus these samples will be mostly confined to a fraction of the RMS value of the large noise at the H side of the HL situation, which is very different from what we have in the LH situation outlined above. In this way we have a heuristic hope that the mean-square values of these voltage samples will depend on the bit situations (HL or LH).

Note, such an attack would not work against the original KLJN scheme as there the HL and LH voltage and resistor pairs are identical.

Moreover, whenever the net power flow is zero due to thermal equilibrium, such as in the original KLJN scheme, the wire voltage and current are uncorrelated. Then their Gaussianity implies that sampling the voltage at the zero-crossing times of the current represents an independent sampling of the voltage. That means the mean-square wire voltage will be the nominal value for the HL/LH situation thus there is no information there for Eve. This is another reason why the original KLJN scheme would be immune against such a zero-crossing attack. Moreover it is an indication that the FCK1VMG-KLJN system, where the power flow is also zero, is also immune against this new attack.

Below we demonstrate by computer simulations that the intuitive expectation turns out to be valid and the VMG-KLJN scheme is leaking information at nonzero-power flow. In the next section, we also show that there is a new KLJN scheme that is secure against the attack even though the 4 resistors are different.

**Computer simulations/verification of the zero-crossing attack**

During the noise generation we used oversampling to produce sufficiently smooth noises to emulate physical noise sources and to detect zero-crossing current evets with sufficient accuracy.

An example for Alice's and Bob's noise voltages and channel current in the VMG-KLJN scheme, at $R_{HA}$ = 46,416 Ω, $R_{LA}$ = 278 Ω, $R_{HB}$ = 278 Ω, $R_{LB}$ = 100 Ω, $T_{HA}$ = 8.0671 x $10^{18}$ K, $T_{LA}$ = 1.3033 x $10^{17}$ K, $T_{HB}$ = 6.2112 x $10^{16}$ K, $T_{LB}$ = 1.1694 x $10^{17}$ K, and B = 500 Hz, is shown in Figure 5. The zero-crossing points of the channel current are the points where Alice's and Bob's noise voltages are equal. At the particular choice of resistances, in the LH case, Alice and Bob have similar noise voltage amplitudes, while in the HL case, Alice's noise voltage amplitudes are much larger than Bob's, thus the zero-crossing points are ultimately determined by Bob's noise voltage.

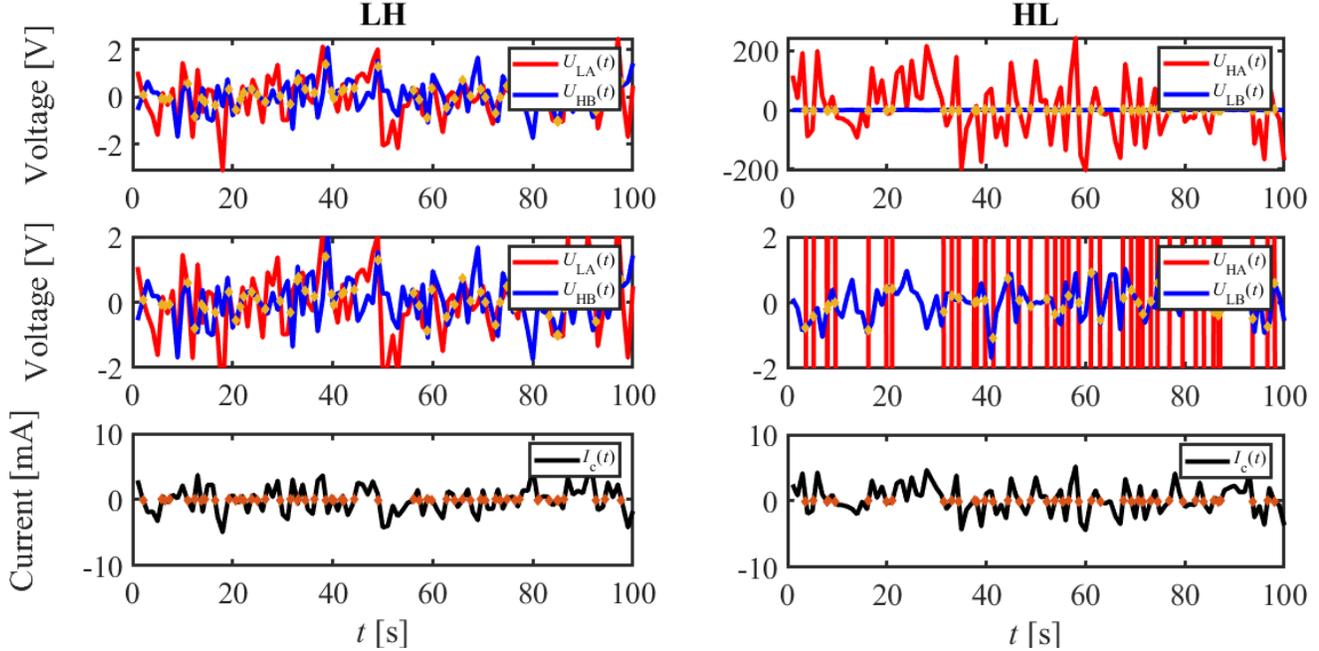

**Figure 5.** A realization of the instantaneous noise voltages of Alice (red) and Bob (blue) and the channel current (black) in the LH (left) and HL (right) cases for the VMG-KLJN scheme, at $R_{HA}$ = 46.4 kΩ, $R_{LA}$ = 278 Ω, $R_{HB}$ = 278 Ω, $R_{LB}$ = 100 Ω, $T_{HA}$ = 8.07 x $10^{18}$ K, $T_{LA}$ = 1.30 x $10^{17}$ K, $T_{HB}$ = 6.21 x $10^{16}$ K, $T_{LB}$ = 1.17 x $10^{17}$ K, and B = 500 Hz. $U_{LA}^2$ = 1 $V^2$, $U_{BH}^2$ = 0.477 $V^2$, $U_{HA}^2$ = 1.03 x $10^4$ $V^2$, and $U_{LB}^2$ = 0.323 $V^2$. The points where the channel current $I_c(t)$ is zero, represented in orange, are the points where Alice's and Bob's noise voltages are equivalent, represented in yellow. In the LH case, Alice's noise voltage $U_{LA}(t)$ is comparable to Bob's noise voltage $U_{HB}(t)$, while in the HL case, Alice's noise voltage $U_{LA}(t)$ is significantly larger than Bob's noise voltage $U_{HB}(t)$, thus the points where Alice's and Bob's noise voltages are equal to each other are ultimately determined by the smaller noise amplitude. $U_{LB} \gg U_{HA}$, thus $U_{LB}(t)$ looks like a straight line because of limited resolution in the figure.

The histograms of the mean-square channel voltages, currents, and zero-crossing points after 1,000 runs are shown in Figure 6 for the original KLJN scheme (a), the VMG-KLJN scheme (b), and the FCK1-VMG-KLJN scheme (c). The orange histograms represent the LH situation, whereas the blue histograms represent the HL situation. The red lines represent the expected (mean) value.

The secure bit (HL and HL) mean-square voltages and currents are the same in all schemes as it has been expected by the VMG-KLJN creators.

However, HL and LH the zero-crossing sampled mean-square voltages $U^2_{c,zc}$ are the same only in the original KLJN and FCK1-VMG-KLJN schemes, but markedly different in the VMG-KLJN scheme indicating its cracked security.

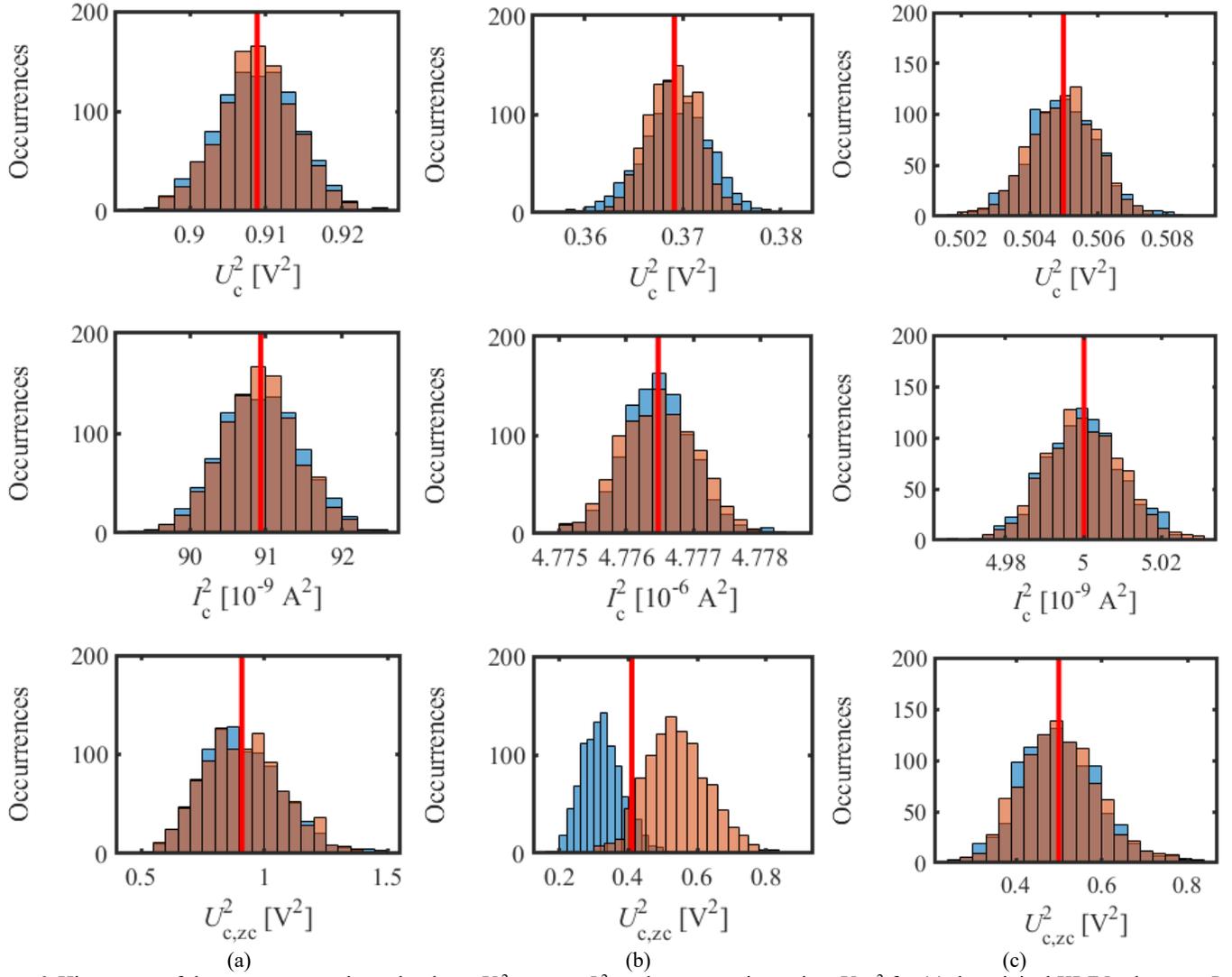

**Figure 6.** Histograms of the mean-square channel voltage $U_c^2$, current $I_c^2$, and zero-crossing points $U_{c,zc}^2$ for (a) the original KLJN scheme at $R_{HA}$ = $R_{HB}$ = 10 kΩ 46.4 kΩ and $R_{LA}$ = $R_{LB}$ = 1 kΩ, (b) the VMG-KLJN scheme at $R_{HA}$ = 46.4 kΩ, $R_{LA}$ = 278 Ω, $R_{HB}$ = 278 Ω, and $R_{LB}$ = 100 Ω, and (c) the FCK1-VMG-KLJN scheme at $R_{HA}$ = 100 kΩ, $R_{LA}$ = 10 kΩ, $R_{HB}$ = 10 kΩ, and $R_{LB}$ = 1 kΩ. The orange histograms represent the LH situation, and the blue histograms represent the HL situation. The red lines represent the expected (mean) value. In all three schemes, $U_c^2$ and $I_c^2$ have the same LH and HL distributions. In the original KLJN and FCK1-VMG-KLJN schemes, $U_{c,zc}^2$ has the same LH and HL distributions, while in the VMG-KLJN scheme, the $U_{c,zc}^2$ LH and HL distributions are dispersed.

Table 1 shows the mean-square voltage $U_c^2$, mean-square current, $I_c^2$, average power $P_{AB}$, and mean-square $U_{c,zc}^2$ of the zero-crossing voltage values for the original KLJN, furthermore for three VMG-KLJN, and three FCK1-VMG-KLJN representations.

In the original KLJN and FCK1-VMG-KLJN representations, the mean-square zero-crossing voltage approaches the channel voltage indicating an random, current-independent sampling.

In the VMG-KLJN scheme, as the cross-correlation between the voltage and current increases (indicated also by the nonzero power flow), the mean-square zero-crossing voltage becomes more dispersed in the LH and HL cases.

Table 2 shows the statistical run for Eve's probability $p$ and its standard deviation $\sigma_p$ of guessing the correct bit. When the average power $P_{AB}$ approaches zero, the $p$ value approaches 0.5 (thus the information leak converges zero) because the cross-correlation coefficient between the current and voltage also converges to zero.

**Table 1.** Results for the wire mean-square voltage $U_c^2$, mean-square current, $I_c^2$, average power $P_{AB}$, and zero-crossing mean-square voltage $U_{c,zc}^2$ for the KLJN, three VMG-KLJN, and FCK1-VMG-KLJN schemes. In the classical KLJN and FCK1-VMG-KLJN schemes, $U_{c,zc}^2$ approaches $U_c^2$. In the VMG-KLJN scheme, as $P_{AB}$ increases, $U_{c,zc}^2$ becomes split in the LH and HL situations.

| Scheme | bit | [a]$R_A$ [Ω] | [b]$R_B$ [Ω] | $U_c^2$ [V²] | $I_c^2$ [$10^{-6}$ A²] | $P_{AB}$ [$10^{-3}$ W] | $U_{c,zc}^2$ [V²] |
|---|---|---|---|---|---|---|---|
| KLJN | LH | 1k | 10k | 0.908 | 0.091 | 0 | 0.907 |
| | HL | 10k | 1k | | | | 0.908 |
| VMG-KLJN | LH | 100 | 16.7k | 0.991 | 0.314 | 0.026 | 0.989 |
| | HL | 16.7k | 278 | | | | 1.009 |
| | LH | 278 | 278 | 0.368 | 4.786 | 0.471 | 0.301 |
| | HL | 46.4k | 100 | | | | 0.576 |
| | LH | 100 | 6k | 0.967 | 0.073 | 0.156 | 0.675 |
| | HL | 360k | 2.2k | | | | 0.845 |
| FCK1-VMG-KLJN | LH | 10k | 10k | 0.500 | 0.005 | 0 | 0.498 |
| | HL | 100k | 1k | | | | 0.502 |

[a]$R_A$ is Alice's choice of resistance. In the HL situation, $R_A = R_{HA}$, and in the LH situation, $R_A = R_{LA}$.
[b]$R_B$ is Bob's choice of resistance. In the HL situation, $R_B = R_{LB}$, and in the LH situation, $R_B = R_{HB}$.

**Table 2.** Statistical run for Eve's probability $p$ of guessing the correct bit from the zero-crossing attack on each scheme. When the average power $P_{AB}$ approaches zero, the $p$ value approaches 0.5 (thus the information leak converges zero) because the cross-correlation coefficient between the current and voltage also converges to zero.

| Scheme | bit | $R_A$ [Ω] | $R_B$ [Ω] | $P_{AB}$ [$10^{-3}$ W] | $p$ | $\sigma_p$ |
|---|---|---|---|---|---|---|
| KLJN | LH | 1k | 10k | 0 | 0.5002 | 0.0091 |
| | HL | 10k | 1k | | | |
| VMG-KLJN | LH | 100 | 16.7k | 0.026 | 0.5885 | 0.0022 |
| | HL | 16.7k | 278 | | | |
| | LH | 278 | 278 | 0.471 | 0.7006 | 0.0053 |
| | HL | 46.4k | 100 | | | |
| | LH | 100 | 6k | 0.156 | 0.6281 | 0.0021 |
| | HL | 360k | 2.2k | | | |
| FCK1-VMG-KLJN | LH | 10k | 10k | 0 | 0.5028 | 0.0091 |
| | HL | 100k | 1k | | | |

In conclusion, the computer simulations confirmed that the zero-crossing attack is an efficient passive attack against the general VMG scheme whenever the net power flow is not zero. The FCK1-VMG-KLJN protocol, which is the zero-power version of the scheme is robust against this attack similarly to the original KLJN scheme.

## CONCLUSION: PERFECT SECURITY REQUIRES EQULIBRIUM

We introduced a new passive attack, which can extract information from the VMG-KLJN system, successfully compromises its security.

On the other hand, the standard KLJN and the FCK1-VMG-KLJN schemes are immune against this attack because their net power flow is zero due their thermal equilibrium feature.

Our results prove that thermal equilibrium is essential for the perfect security of KLJN schemes, including their VMG-KLJN variations. Therefore, the Second Law of Thermodynamics is the fundamental component of the security of the VMG-KLJN system, too.

Nevertheless, we believe that, with careful design and proper compromises, the VMG-KLJN scheme[46] has strong potential for applications in chip technology even if its security level is reduced compared to the features of original KLJN protocol. It is a security level that is sufficient for many practical applications but is reduced due to the deviation from the original KLJN. Its FCK1-VMG-KLN version[58] virtually eliminates the information leak however then only 3 resistors can be freely chosen.